\documentclass[prx,10pt,a4paper,twocolumn,nofootinbib, superscriptaddress]{revtex4-1}
\usepackage{graphicx}
\usepackage{amsmath}
\usepackage{amsfonts}
\usepackage{amssymb}
\usepackage{soul}
\usepackage{dsfont}
\usepackage{hyperref}
\hypersetup{colorlinks=true, citecolor=blue, urlcolor=blue, linkcolor=blue}
\usepackage{amstext}
\usepackage[caption=false]{subfig}
\usepackage{epstopdf} 
\usepackage{mathtools} 

\usepackage{braket}

\newcommand{\RN}[1]{\textup{\uppercase\expandafter{\romannumeral#1}}}

\date{\today}

\begin{document}

\title{Time evolution of two harmonic oscillators with cross-Kerr interactions}

\author{David Edward Bruschi}\email{david.edward.bruschi@gmail.com}
\affiliation{Central European Institute of Technology (CEITEC), Brno University of Technology, Purky\u{n}ova 123, 621 00 Brno, Czech Republic}
\affiliation{Institute for Quantum Optics and Quantum Information - IQOQI Vienna, Austrian Academy of Sciences, Boltzmanngasse 3, 1090 Vienna, Austria}

\begin{abstract}
We study the time evolution of two coupled quantum harmonic oscillators interacting through nonlinear optomechanical-like Hamiltonians that include cross-Kerr interactions. We employ techniques developed to decouple the time-evolution operator and obtain the analytical solution for the time evolution of the system. We apply these results to obtain explicit expressions of a few quantities of interest. Our results do not require approximations and therefore allow to study the nature and implications of the full nonlinearity of the system. Potential applications and extensions to optomechanics are also discussed.
\end{abstract}

\maketitle

\section{Introduction}\label{intro}
The understanding of the dynamics of quantum systems is paramount to deepen our understanding of the the laws of Nature. Given an arbitrary Hamiltonian, it is in general impossible to obtain a full and analytical expression for the time evolution of the system it describes. Typically, approximations or ad-hoc methods are required, leaving open the question of the existence of a general framework or set of tools to address the problem systematically.

In recent years, specific techniques have been developed to tackle the problem for Gaussian states of bosonic systems interacting through quadratic Hamiltonians \cite{puri2001mathematical,bruschi2013time}. This has led to the development of a mathematical framework aimed at obtaining the full time evolution of the system using the power of the covariance matrix formalism \cite{adesso2014continuous}. Interestingly, these techniques do not apply to this class of states and transformations only, but it can be generalised to arbitrary states and arbitrary Hamiltonians. This novel approach has led to very interesting preliminary results, where complete and analytical solutions of the time evolution of the system were found for optomechanical-like Hamiltonians of arbitrary numbers of modes of light interacting with arbitrary numbers of mechanical resonators \cite{Bruschi:Xuereb:2018,Bruschi:2018}. This has allowed, for example, for a quantitative analysis of the nonlinear character of an optomechanical system \cite{qvarfort2018enhanced}.

In this work we take another step forward and study the time evolution of a system of two bosonic modes interaction via an optomechanical-like Hamiltonian with cross Kerr terms. Such Hamiltonian does not model a realistic optomechanical systems since it does not take into account external drive, loss or dissipation. However, this Hamiltonian models ideal quantum systems interacting through nonlinear terms that can potentially be designed in realistic implementations where loss and decoherence can be ignored at first. We are able to find an analytical solution to the time evolution which we use to compute the expectation value of relevant quantities as a function of time,  such as the number of `phonons' in the system, as well as the mixedness induced by the interaction in the state of the resonator's subsystem. We specialise our results to a few scenarios of interest, which model light-matter interaction that can be exploited for sensing and gravimentry \cite{Raetzel:Schneiter:2018}, as well as tests of fundamental physics \cite{Schmoele:Dragostis:2016}.

Ultimately, our work can aid in deepening our understanding of the mathematical framework of quantum mechanics, one of our two pillars of modern science.

The paper is structured as followed: In Section \ref{general:framework} we introduce the general framework necessary for our work. In Section \ref{decoupling:section} we decouple the time evolution of the system for the general case. In Section \ref{applications:section} we specialise to scenarios of more practical interest. Finally, we briefly give concluding remarks in Section \ref{conclusions}. 

\section{General framework}\label{general:framework}
We start by introducing the Hamiltonian and the tools necessary to this work. More details about the computations are left to the appendices.

\subsection{Hamiltonian}\label{optomech}
In this work we consider the general Hamiltonian $ \hat {H} =  \hat{H}_0(t)+\hat{H}_\textrm{I}(t)$ where $\hat{H}_0:=\hbar\,\omega_\mathrm{c}(t) \hat a^\dagger \hat a + \hbar\,\omega_\mathrm{m} \,\hat b^\dagger \hat b$ is the \textit{potentially time-dependent} free Hamiltonian and
\begin{align}\label{main:time:dependent:Hamiltonian:to:decouple:dimensionful:general} 
\hat{H}_\textrm{I}(t) =& \hbar\,g^{(+)}_{1}(t)\,\hat a^\dagger\hat a \,\hat{B}_++ \hbar\,g^{(-)}_{1}(t)\,\hat a^\dagger\hat a \,\hat{B}_-\nonumber\\
 &  +  \hbar \, g^{(+)}_2(t)\,\hat a^\dagger\hat a \,\hat{B}^{(2)}_++  \hbar \, g^{(-)}_2(t)\,\hat a^\dagger\hat a \,\hat{B}^{(2)}_-\nonumber\\
 &+2\,\hbar \, g_2^\prime(t)\,\hat a^\dagger\hat a\,\hat b{}^\dagger \hat b,
\end{align}
we have defined $\hat{B}_+:=\hat b^{\dagger}+\hat b$, $\hat{B}_-:=i(\hat b^{\dagger}-\hat b)$, $\hat{B}^{(2)}_+:=\hat b^{\dagger2}+\hat b^2$ and $\hat{B}^{(2)}_-:=i(\hat b^{\dagger2}-\hat b^2)$ for notational convenience. These operators correspond to the quadrature operators, and the square of the quadrature operators, respectively. We choose to retain our notation because it is more natural to the techniques used in this work.
An advantage of the expression \eqref{main:time:dependent:Hamiltonian:to:decouple:dimensionful:general} is that it allows us to consider each cross-Kerr term independently from each other. 

The Hamiltonian \eqref{main:time:dependent:Hamiltonian:to:decouple:dimensionful:general} is a formal extension of well-known Hamiltonians that can model light and matter interaction, e.g., optomechanical Hamiltonians. The main difference is that we consider an ``ideal'' case here, namely, we do not include an external (laser) drive and we assume that the system is lossless and noiseless. In this sense, the system considered here does not model realistic optomechanical systems. Nevertheless, it provides a platform for studying cross-Kerr interactions, which can become of practical interest with the advance of technological control.

%
%

\subsection{Dimensionless dynamics}\label{optomech}
To understand which are the relevant dimensionless parameters that govern the dynamics of the system we introduce dimensionless quantities and rescale the Hamiltonian \eqref{main:time:dependent:Hamiltonian:to:decouple:dimensionful:general}. This is achieved rescaling all frequencies and time by $\omega_{\mathrm{m}}$. Therefore, the laboratory time $t$ will become $\tau=\omega_{\mathrm{m}}\,t$, where $\tau$ is the new dimensionless time. The couplings in the Hamiltonian are subsequently relabelled as follows: $\tilde{g}^{(\pm)}_1(\tau) = g^{(\pm)}_1(\omega_{\mathrm{m}}\,t ) / \omega_{\mathrm{m}}$, $\tilde{g}^{(\pm)}_2(\tau) = g^{(\pm)}_2(t \omega_{\mathrm{m}})/\omega_{\mathrm{m}}$ and $\tilde{g}^\prime_2(\tau) = g^\prime_2(t \omega_{\mathrm{m}})/\omega_{\mathrm{m}}$. We also rescale the optical frequency to $\tilde{\omega}_\mathrm{c}(\tau):=\omega_\mathrm{c}(\omega_{\mathrm{m}}\,t)/\omega_\mathrm{m}$ and the Hamiltonian by $\hbar$. This implies that we use the rescaled Hamiltonian $ \hat {H}(\tau) =  \hat{H}_0(\tau)+\hat{H}_\textrm{I}(\tau)$, where we have $\hat{H}_0(\tau):=\tilde{\omega}_\mathrm{c}(\tau)\,\hat a^\dagger \hat a + \hat b{}^\dagger \hat b$ and 
\begin{align}\label{main:time:dependent:Hamiltonian:to:decouple:general}
	\hat {H}_\textrm{I}(\tau) =&\tilde{g}^{(+)}_{1}(\tau)\,\hat a^\dagger\hat a \,\hat{B}_++ \tilde{g}^{(-)}_{1}(\tau)\,\hat a^\dagger\hat a \,\hat{B}_-+2\,\tilde{g}_2^\prime(\tau)\,\hat a^\dagger\hat a\,\hat b{}^\dagger \hat b\nonumber\\
 &  +  \tilde{g}^{(+)}_2(\tau)\,\hat a^\dagger\hat a \,\hat{B}^{(2)}_++  \tilde{g}^{(-)}_2(\tau)\,\hat a^\dagger\hat a \,\hat{B}^{(2)}_-.
 \end{align}

\subsection{Solving the dynamics}\label{tools}
Here we outline the tools needed to solve the dynamics generated by \eqref{main:time:dependent:Hamiltonian:to:decouple:general}. We refer the reader to Appendices \ref{appendix:decoupling} and \ref{appendix:Hamiltonian:nonlinear:decoupling} for detailed calculations.

The time-evolution operator of a quantum system with time dependent Hamiltonian $\hat{H}(t)$ reads 
\begin{equation}\label{general:time:evolution:operator}
\hat{U}(t)=\overset{\leftarrow}{\mathcal{T}}\,\exp\left[-\frac{i}{\hbar}\int_0^{t} dt'\,\hat{H}(t')\right],
\end{equation}
where $\overset{\leftarrow}{\mathcal{T}}$ is the time ordering operator \cite{bruschi2013time}.\footnote{An alternative approach was attempted in \cite{Brown:Martin-Martinez:2013}} This expression simplifies to $U(t)=\exp[-\frac{i}{\hbar}\,\hat{H}\,t]$ when the Hamiltonian $\hat{H}$ is time independent. However, we are interested in Hamiltonians with time dependent parameters.
 
Any Hamiltonian can be cast in the form $\hat{H}(t)=\sum_n \hbar\, g_n(t)\,\hat{G}_n$, where the $\hat{G}_n$ are time-independent Hermitian operators, the $g_n(t)$ are time-dependent real functions, and the choice of $\hat{G}_n$ is not unique.

In general, one aims to recast \eqref{general:time:evolution:operator} in the form
\begin{equation}\label{decoupled:time:evolution:operator}
\hat{U}(t)=\prod_n\hat{U}_n(t),
\end{equation}
where we have defined $\hat{U}_n:=\exp[-i\,F_n(t)\,\hat{G}_n]$ and the real functions $F_n(t)$ are in general time dependent. If  such an expression exists, we say that the time-evolution operator has been \textit{decoupled}. The functions $F_n(t)$ can be found using the techniques developed in the literature \cite{bruschi2013time} and are determined solely by the parameters of the Hamiltonian. The order of the operators in \eqref{decoupled:time:evolution:operator} is not unique. A different order will change the form of the functions $F_n(t)$ but not the expectation value of measurable quantities. A more detailed outline of these techniques can be found in Appendix \ref{appendix:decoupling}.

\subsection{Initial state}
The last ingredient in this work is the choice of the initial state of the system. We assume that the optical mode is initially in a coherent state $\ket{\mu_{\mathrm{c}}}$, while the mechanical is in a thermal state $\hat{\rho}_{\mathrm{m}}(T)$. These states are defined by $\hat{a} \ket{\mu_{\mathrm{c}}} = \mu_{\mathrm{c}} \ket{\mu_{\mathrm{c}}}$ and $\hat{\rho}_{\mathrm{m}}(T)=\sum_n \frac{\tanh^{2n} r}{\cosh^2 r} \ket{n}\bra{n}$, and $\tanh r:=\exp[-\frac{\hbar\,\omega_{\mathrm{m}}}{2\,k_{\mathrm{B}}\,T}]$. 
 
Therefore, the initial state $|\psi(0)\rangle$  reads
\begin{align}\label{initial:state:two}
\hat{\rho}(0) =\ket{\mu_{\mathrm{c}}}\bra{\mu_{\mathrm{c}}}\otimes \hat{\rho}_{\mathrm{m}}(T),
\end{align}
where $N_\text{a}(0)=|\mu|^2$ and $N_\text{b}(0)=\sinh^2r$ are the initial number of excitation of modes $\hat{a}$ and $\hat{b}$ respectively.
This state is a good approximation for the initial state of realistic systems (i.e., optomechanical systems).

\section{Decoupling of a non-linear time-dependent optomechanical cross Kerr Hamiltonian }\label{decoupling:section}
The first aim of this work is to obtains an analytical solution to the decoupled time-evolution operator \eqref{decoupled:time:evolution:operator} given our Hamiltonian \eqref{main:time:dependent:Hamiltonian:to:decouple:general}. Techniques exist already that can be used to tackle to decoupling of time evolution operators in the way that is suitable for us \cite{bruschi2013time}. These are the techniques we will employ here.

\subsection{Decoupling algebra of the nonlinear Hamiltonian}
Decoupling of the Hamiltonian \eqref{main:time:dependent:Hamiltonian:to:decouple:general} can be done using a choice of the Hermitian operators $\hat{G}_n$, see Appendix \ref{appendix:decoupling}.

The first step is to write
\begin{align}
\hat U(\tau)=&e^{-i\,\int_0^\tau\,d\tau'\tilde{\omega}_\textrm{c}(\tau') \hat{a}^\dag\,\hat{a}}\,e^{-i\,\hat{\theta}_2\,\hat{b}^\dag\,\hat{b}}\,\hat U_\textrm{I}(\tau),
\end{align}
where we have introduced $\hat{\theta}_2=\hat{\theta}_2(\hat{a}^\dag\,\hat{a}):=\tau+2\,\int_0^\tau\,d\tau'\,\tilde{g}_2^\prime(\tau')\,\hat a^\dagger\hat a$ for notational convenience. Notice that $\hat{\theta}_2$ is a time-dependent operator-function of the Hermitian operator $\hat{a}^\dag\,\hat{a}$. We can think of $\hat U_\textrm{I}(\tau)$ as the time evolution operator of the interaction picture. 

We now need to consider the operator $\hat U_\textrm{I}(\tau):=\overset{\leftarrow}{\mathcal{T}}\exp\bigl[-\frac{i}{\hbar}\int_0^\tau \hat{H}_\textrm{I}(\tau)\bigr]$ induced by the Hamiltonian $\hat{H}_\textrm{I}(\tau)=\hat{H}_{\mathrm{sq}}(\tau)+\hat{H}_{\mathrm{OM}}(\tau)$, where
\begin{align}
\hat{H}_{\mathrm{sq}}(\tau)&:=a^\dagger\hat a\,\left[\tilde{g}^{(+)}_2(\tau)\,\cos(2\,\hat{\theta}_2)-\tilde{g}^{(-)}_2(\tau)\,\sin(2\,\hat{\theta}_2)\right]\,\hat{B}^{(2)}_+\nonumber\\
	&+a^\dagger\hat a\,\left[\tilde{g}^{(+)}_2(\tau)\,\sin(2\,\hat{\theta}_2)+\tilde{g}^{(-)}_2(\tau)\,\cos(2\,\hat{\theta}_2)\right]\,\hat{B}^{(2)}_-\nonumber\\
	\hat{H}_{\mathrm{OM}}(\tau)&:=a^\dagger\hat a\,\left[\tilde{g}^{(+)}_1(\tau)\,\cos(\hat{\theta}_2)-\tilde{g}^{(-)}_1(\tau)\,\sin(\hat{\theta}_2)\right]\,\hat{B}_+\nonumber\\
	&+a^\dagger\hat a\,\left[\tilde{g}^{(+)}_1(\tau)\,\sin(\hat{\theta}_2)+\tilde{g}^{(-)}_1(\tau)\,\cos(\hat{\theta}_2)\right]\,\hat{B}_-.
\end{align}
We then split the evolution as $\hat U_\textrm{I}(\tau)=\hat{U}_{\mathrm{sq}}(\tau)\,\hat{U}_{\mathrm{OM}}(\tau)$, where we have defined
\begin{align}
\hat{U}_{\mathrm{sq}}:=&\overset{\leftarrow}{\mathcal{T}}\exp\left[-\frac{i}{\hbar}\int_0^\tau\,d\tau'\,\hat{H}_{\mathrm{sq}}(\tau')\right]\nonumber\\
\hat{U}_{\mathrm{OM}}:=&\overset{\leftarrow}{\mathcal{T}}\exp\left[-\frac{i}{\hbar}\int_0^\tau\,d\tau'\,\hat{U}^\dag_{\mathrm{sq}}\,\hat{H}_{\mathrm{OM}}(\tau')\,\hat{U}_{\mathrm{sq}}\right].
\end{align}
We need to supplement this ansatz with a second one, namely with the action of the two-mode squeezing-like operator $\hat{U}_{\mathrm{sq}}$ on the operators $\hat{b}$ and $\hat{b}^\dag$. We have
\begin{align}\label{two:mode:squeezing}
\hat{U}_{\mathrm{sq}}^\dag\,\hat{b}\,\hat{U}_{\mathrm{sq}}=&\hat{\alpha}\,\hat{b}+\hat{\beta}\,\hat{b}^\dag,
\end{align}
where $\hat{\alpha}=\hat{\alpha}(\hat{a}^\dagger\hat a)$ and $\hat{\beta}=\hat{\beta}(\hat{a}^\dagger\hat a)$ are the Bogoliubov coefficients. These coefficients are \textit{time-dependent} functions of the operator $\hat{a}^\dagger\hat a$. They satisfy the Bogoliubov identities $\hat{\alpha}\,\hat{\alpha}^\dag-\hat{\beta}\,\hat{\beta}^\dag=\mathds{1}$ and $\hat{\alpha}\,\hat{\beta}^\dag-\hat{\beta}\,\hat{\alpha}^\dag=0$.

This allows us to obtain
\begin{align}
\hat{U}_{\mathrm{sq}}^\dag\,\hat{H}_{\mathrm{OM}}(\tau)\,\hat{U}_{\mathrm{sq}}=&\hat{a}^\dagger\hat a\,\hat \Re\left[\tilde{g}_1(\tau)\,\hat{E}_2\,(\hat{\alpha}^\dag+\hat{\beta}^\dag)\right]\,\hat{B}_+\nonumber\\
&-i\,\hat{a}^\dagger\hat a\,\hat \Im\left[\tilde{g}_1(\tau)\,\hat{E}_2\,(\hat{\alpha}^\dag-\hat{\beta}^\dag)\right]\,\hat{B}_-,
\end{align}
where we have defined $\hat{E}_2=\hat{E}_2(\hat{a}^\dagger\hat a):=\cos(\hat{\theta}_2)+i\,\sin(\hat{\theta}_2)$, the coupling $\tilde{g}_1(\tau):=\tilde{g}^{(+)}_1(\tau)+i\,\tilde{g}_1^{(-)}(\tau)$ and $\Re(\hat{O}):=\frac{1}{2}[\hat{O}+\hat{O}^\dag]$ for convenience of notation. 

Given all of the above, we can finally obtain
\begin{align}\label{U}
\hat U(t):=&e^{-i\,\int_0^\tau\,d\tau'\tilde{\omega}_\textrm{c}(\tau') \hat{a}^\dag\,\hat{a}+i\,\hat{F}^{(2)}\,(\hat{a}^\dagger\hat a\,)^2}\, e^{-i\,\hat{\theta}_2\,\hat{b}^\dag\,\hat{b}}\,\hat{U}_{\mathrm{sq}}\nonumber\\
&\times e^{-i\,\int_0^\tau\,d\tau'\,\hat{a}^\dagger\hat a\,\Re\left[\tilde{g}_1(\tau')\,\hat{E}_2\,(\hat{\alpha}^\dag+\hat{\beta}^\dag)\right]\,\hat{B}_+}\nonumber\\
&\times e^{i\,\int_0^\tau\,d\tau'\,\hat{a}^\dagger\hat a\,\Im\left[\tilde{g}_1(\tau')\,\hat{E}_2\,(\hat{\alpha}^\dag-\hat{\beta}^\dag)\right]\,\hat{B}_-},
\end{align}
which is the main expression for the decoupled  time-evolution operator in this work.

To complete our main result \eqref{U} we require the expression of $\hat{F}^{(2)}$ and of the Bogoliubov coefficients $\hat{\alpha}$ and $\hat{\beta}$. The expression for $\hat{F}^{(2)}$ is readily found as
\begin{align}\label{F2:general}
\hat{F}^{(2)}=& 2\,\int_0^\tau\,d\tau'\,\Im\left[\tilde{g}_1(\tau')\,\hat{E}_2\,(\hat{\alpha}^\dag-\hat{\beta}^\dag)\right]\nonumber\\
&\times\int_0^{\tau'}\,d\tau''\,\Re\left[\tilde{g}_1(\tau'')\,\hat{E}_2\,(\hat{\alpha}^\dag+\hat{\beta}^\dag)\right].
\end{align}
This leaves us with the task of computing the Bogoliubov coefficients $\hat{\alpha}$ and $\hat{\beta}$ in order to obtain a fully analytical understanding of our system.

\subsection{The action of the single-mode squeezing operator}
We have noted that, in general, the action of $\hat{U}_{\mathrm{sq}}$ on the operators $\hat{b}$ has the form \eqref{two:mode:squeezing}. Ideally, we would like to have an analytical expression for the functional form of $\hat{\alpha}$ and $\hat{\beta}$ in terms of $\hat{a}^\dagger\hat a$, and of the couplings of the system. Although this is not possible in general \cite{moore2016tuneable}, we proceed to construct two uncoupled differential equations that relate the derivatives of $\hat{\alpha}$ and $\hat{\beta}$ with the couplings $\tilde{g}^{\pm}_2(\tau)$ and $\tilde{g}_2^\prime(\tau)$. 

The expression for the single mode squeezing is given in \eqref{two:mode:squeezing}.
In Appendix \ref{appendix:Hamiltonian:nonlinear:decoupling} we show that
\begin{align}\label{Bogoliubov:coefficients:main:text}
\hat{\alpha}=&\hat{p}_{11} \nonumber\\
\hat{\beta}=&-i\,a^\dagger\hat a\, \int_0^\tau d\tau'\,\tilde{\chi}_2(\tau')\,e^{2\,i\,(\phi_2+\hat{\theta}_2)}\,\hat{p}^\dag_{11},
\end{align}
where we have introduced the modulus $\tilde{\chi}(\tau):=\sqrt{\tilde{g}^{(+)2}_2(\tau)+\tilde{g}^{(-)2}_2(\tau)}$, the angle $\phi_2$ through $\tan(2\,\phi_2)=\tilde{g}^{(-)}_2/\tilde{g}^{(+)}_2$ and the operator-function $\hat{p}_{11}=\hat{p}_{11}\left(\int_0^\tau\,d\tau'\,\tilde{\chi}(\tau')\right)$, which satisfies the second-order differential equation
\begin{align}\label{main:differential:equation:text}
\tilde{\chi}_2\,\ddot{\hat{p}}_{11}+2\,i\,\frac{d}{d\tau}(\phi_2+\hat{\theta}_2)\,\dot{\hat{p}}_{11}-\tilde{\chi}_2\,(a^\dagger\hat a)^2\,\hat{p}_{11}=&0.
\end{align} 
The derivative here is with respect to $y(\tau):=\int_0^\tau\,d\tau'\tilde{\chi}_2(\tau')$.
These differential equations have to be supplemented by the initial conditions $\hat{p}_{11}(0)=\mathds{1}$ and $\dot{\hat{p}}_{11}(0)=0$. 

The evolution of the initial state $\hat{\rho}(0)$ is obtained by the usual Heisenberg equation $\hat{\rho}(\tau)=\hat{U}(\tau)\,\hat{\rho}(0)\,\hat{U}^\dag(\tau)$. The full expression is not illuminating and we do not print it here. However, we can ask what is the expression for the time evolution of the mode operators $\hat{a}$ and $\hat{b}$. These expressions allow us, in principle, to compute the expectation value of most quantities of interest.

We define $\hat{a}(\tau):=\hat{U}^\dag(\tau)\,\hat{a}\,\hat{U}(\tau)$ and $\hat{b}(\tau):=\hat{U}^\dag(\tau)\,\hat{b}\,\hat{U}(\tau)$ and, using the expression \eqref{U}, we find
\begin{align}\label{first:moment:mechanical:mode:main}
\hat{b}(\tau)=&\hat{E}^\dag_2\left\{\hat{\alpha}\,\hat{b}+\hat{\beta}\,\hat{b}^\dag-i\,a^\dagger\hat a\,\hat{\alpha}\,\hat{I}+i\,a^\dagger\hat a\,\hat{\beta}\,\hat{I}^\dag\right\}.
\end{align}
Here $\hat{I}=\hat{I}(\hat{a}^\dag\,\hat{a}):=\int_0^\tau\,d\tau'\left[\tilde{g}_1(\tau')\,\hat{E}_2\,\hat{\alpha}^\dag+\tilde{g}_1^*(\tau')\,\hat{E}^\dag_2\,\hat{\beta}\right]$ for notational convenience.
The expression for $\hat{a}(\tau)$ depends on the explicit functional form of $\hat{F}^{(2)}$, $\Im\bigl[\tilde{g}_1(\tau')\,\hat{E}_2\,(\hat{\alpha}^\dag-\hat{\beta}^\dag)\bigr]$ and $\Re\bigl[\tilde{g}_1(\tau'')\,\hat{E}_2\,(\hat{\alpha}^\dag+\hat{\beta}^\dag)\bigr]$ in terms of $a^\dagger\hat a$. This can be computed once the scenario of interest has been determined.

The number $N_\text{b}(\tau):=\langle\hat{U}^\dag(\tau)\,\hat{b}^\dag\,\hat{b}\,\hat{U}(\tau)\rangle$ of `phonons' at any time $\tau$ can be computed using \eqref{first:moment:mechanical:mode:main}, and it reads
\begin{align}\label{number:operator:expectation:value:mechanical:mode:main}
N_\text{b}(\tau)=&e^{-|\mu|^2}\sum_n\,\frac{|\mu|^{2\,n}}{n!}\left[(1+2\,|\beta_n|^2)\,(N_\text{b}(0)+n^2\,|I_n\,|^2)\right.\nonumber\\
&\left.+|\beta_n|^2-2\,n^2\,\Re\left(\alpha_n\,\beta^*_n\,I_n^2\right)\right].
\end{align}
In the expression of \eqref{number:operator:expectation:value:mechanical:mode:main} the subscript $n$ means within each operator we need to replace $n\rightarrow\hat{a}^\dag\,\hat{a}$.

\section{Applications: time evolution of specific systems with cross-Kerr terms} \label{applications:section}
We are now in the position to study the time evolution induced by the Hamiltonian \eqref{U} within some specific cross-Kerr scenarios. We will be able to obtain some analytical expressions for meaningful quantities that encode the full nonlinear character of the system.

\subsection{Cross-Kerr without squeezing}
Here we start with the scenariowhere $\tilde{g}^{(\pm)}_2(\tau)=\tilde{\chi}_2(\tau)=0$. This implies that the only cross Kerr term that does \textit{not} vanish is the term $\tilde{g}_2^\prime(\tau)\,\hat a^\dagger\hat a\,\hat b{}^\dagger \hat b$ in \eqref{main:time:dependent:Hamiltonian:to:decouple:general}.
Since $\tilde{\chi}_2(\tau)=0$, we can immediately see that $\hat{p}_{11}=\mathds{1}$ and therefore $\hat{\alpha}=\mathds{1}$ and $\hat{\beta}=0$. Some algebra allows us then to find the expression for \eqref{U} in this case, which reads 
\begin{align}\label{U:simple:cross:Kerr}
\hat U(t)&=e^{-i\,\int_0^\tau\,d\tau'\tilde{\omega}_\textrm{c}(\tau') \hat{a}^\dag\,\hat{a}+i\,\hat{F}^{(2)}\,(\hat{a}^\dagger\hat a\,)^2}\, e^{-i\,\hat{\theta}_2\,\hat{b}^\dag\,\hat{b}}\nonumber\\
&\times e^{-i\,\int_0^\tau\,d\tau'\,\hat{a}^\dagger\hat a\,\left[\tilde{g}^{(+)}_1(\tau)\,\cos(\hat{\theta}_2)-\tilde{g}^{(-)}_1(\tau)\,\sin(\hat{\theta}_2)\right]\,\hat{B}_+}\nonumber\\
&\times e^{i\,\int_0^\tau\,d\tau'\,\hat{a}^\dagger\hat a\,\left[\tilde{g}^{(+)}_1(\tau)\,\sin(\hat{\theta}_2)+\tilde{g}^{(-)}_1(\tau)\,\cos(\hat{\theta}_2)\right]\,\hat{B}_-}.
\end{align}
where $\hat{F}^{(2)}$ reads
\begin{align}\label{F2}
\hat{F}^{(2)}=& 2\,\int_0^\tau\,d\tau'\,\left[\tilde{g}^{(+)}_1(\tau')\,\sin(\hat{\theta}_2)+\tilde{g}^{(-)}_1(\tau')\,\cos(\hat{\theta}_2)\right]\nonumber\\
&\times\int_0^{\tau'}\,d\tau''\,\left[\tilde{g}^{(+)}_1(\tau'')\,\cos(\hat{\theta}_2)-\tilde{g}^{(-)}_1(\tau'')\,\sin(\hat{\theta}_2)\right].
\end{align}
Notice that, for $\tilde{g}_2'(\tau)=\tilde{g}^{(-)}_1=0$, we recover the results found in the literature, as expected \cite{Bruschi:Xuereb:2018}.

The expressions \eqref{U:simple:cross:Kerr} allow us to find 
\begin{align}\label{mode:operator:time:evolution:simple:cross:Kerr}
\hat{b}(\tau)=&e^{-i\,\hat{\theta}_2}\,\left[\hat{b}-i\,\hat a^\dagger\hat a\,\int_0^\tau\,d\tau'\,\tilde{g}_1(\tau')\,e^{i\,\hat{\theta}_2}\right],
\end{align}
where we have omitted the expression for $\hat{a}(\tau)$ which \textit{can} be computed but is not illuminating.

The change $\Delta N_\text{b}(\tau):=N_{\text{b}}(\tau)-N_{\text{b}}(0)$ in the number of phonons, given our initial state \eqref{initial:state:two}, reads
\begin{align}\label{phonon:number:operator:time:evolution:simple:cross:Kerr}
\Delta N_\text{b}(\tau)=&|\mu|^2\,e^{-|\mu|^2}\sum_n\,\frac{(n+1)\,|\mu|^{2\,n}}{n!}\,\left|I_{n+1}\right|^2,
\end{align}
where $I_n=\int_0^\tau\,d\tau'\tilde{g}_1(\tau')\,\exp[i\,(\tau+2\,n\,\int_0^\tau\,d\tau'\,\tilde{g}_2^\prime(\tau'))]$ for this case. 

Finally, we can also compute the mixedness of the reduced state. The calculations can be found in Appendix~\ref{appendix:subsection:mixedness:computation}. 
Lengthy algebra, and the use of a similar approach developed in the literature \cite{Bruschi:2018}, allow us to find
\begin{align}\label{linear:entropy:final:text}
S_N(\hat{\rho}_{\textrm{m}}(\tau))=&1-e^{-2\,|\mu|^2}\sum_{n,n'}\frac{|\mu|^{2\,n+2\,n'}}{n!\,n'!}\,\frac{\exp\left[-\frac{\left|\Delta_{nn'}\right|^2}{\cosh r_T}\,\right]}{\cosh r_T},
\end{align}
where we have defined $\Delta_{nn'}:=\int_0^\tau\,d\tau'\,\tilde{g}_1(\tau')\bigl(n\,e^{i\,\theta_{2,n}(\tau')}-n'\,e^{\theta_{2,n'}(\tau')}\bigr)$ and $\theta_{2,n}(\tau):=\tau+2\,i\,n\,\int_0^{\tau}\,d\tau''\,\tilde{g}_2^\prime(\tau'')$ for convenience of presentation. It is easy to find the result for zero temperature: it is sufficient to set $r_T=0$ in \eqref{linear:entropy:final:text}. Notice that when $\Delta_{nn'}=0$, i.e., $\tilde{g}_1(\tau)=0$, we obtain $S_N(\hat{\rho}_{\textrm{m}}(\tau))=S_N(\hat{\rho}_{\textrm{m}}(0))$ as expected.

\subsection{Cross-Kerr with squeezing and without diagonal term}
Here we consider the simpler scenario when $\tilde{g}_2'(\tau)=0$. This implies that the cross Kerr term that \textit{vanishes} is $\tilde{g}_2^\prime(\tau)\,\hat a^\dagger\hat a\,\hat b{}^\dagger \hat b$ in \eqref{U}.
Since $\tilde{g}_2'(\tau)=0$, the main differential equation \eqref{main:differential:equation:text} reads
\begin{align}
\tilde{\chi}_2(\tau)\,\ddot{p}_{11}-i\,\left(1+\frac{d}{d\tau}\phi_2\right)\,\dot{p}_{11}-\tilde{\chi}_2(\tau)\,(a^\dagger\hat a)^2\,\hat{p}_{11}=&0.
\end{align}
This case, requires numerical integration of the differential equation above, since there is no analytical solution for a general form of the coupling $\tilde{\chi}_2(\tau)$.
A numerical approach can then enable the used of \eqref{U}.

\subsection{Cross-Kerr with constant squeezing and constant diagonal term}
In this final case we consider $\tilde{g}_2'(\tau)=\tilde{g}_2'$ and $\tilde{g}_2(\tau)=\tilde{g}_2$ constant. This implies that also $\tilde{\chi}_2$ and $\phi_2$ are constant. In turn, this means that  \eqref{main:differential:equation:text} reads
\begin{align}
\tilde{\chi}_2\,\ddot{p}_{11}-i\,(1+2\,\tilde{g}_2'\,\hat a^\dagger\hat a)\,\dot{p}_{11}-\tilde{\chi}_2\,(a^\dagger\hat a)^2\,\hat{p}_{11}=&0,
\end{align}
given that $\hat{\theta}_2=(1+2\,\tilde{g}_2'\,\hat a^\dagger\hat a)\,\tau$ in this scenario.
This allows us to find the solution
\begin{align}\label{Bogoliubov:coefficient:all:constant:case}
\hat{p}_{11}(\tau)=&e^{i\,\hat{K}\,\tau}\,\left[\cos\left(\hat{\Lambda}\,\tau\right)-i\,\frac{\hat{K}}{\hat{\Lambda}}\,\sin\left(\hat{\Lambda}\,\tau\right)\right],
\end{align}
where we have introduced $\hat{K}:=1+2\,\tilde{g}_2'\,\hat a^\dagger\hat a$ and $\hat{\Lambda}:=\sqrt{\hat{K}^2-4\,\tilde{\chi}_2^2\,(a^\dagger\hat a)^2}$ for convenience of presentation. Note that $\hat{\Lambda}^{-1}\,\sin(\hat{\Lambda}\,\tau)$ is a well defined operator, since it is the short hand notation for the expression $\hat{\Lambda}^{-1}\,\sin\bigl(\hat{\Lambda}\,\tau\bigr)=\bigl[1-(\hat{\Lambda}\,\tau)^2/3!+(\hat{\Lambda}\,\tau)^4/5!-\ldots\bigr]\,\tau$.
Also note that $\hat{\theta}_2=\hat{K}\,\tau$ in our present case. Finally, note that if we were to set $\tilde{\chi}_2=0$, we would have $\hat{\Lambda}=\hat{K}$ and we can immediately see that $\hat{p}_{11}(\tau)=\mathds{1}$, as expected from our results above.

Recalling that $\hat{p}_{11}(\tau)$ in \eqref{Bogoliubov:coefficient:all:constant:case} gives us one of the Bogoliubov coefficients directly, i.e., $\hat{\alpha}=\hat{p}_{11}(\tau)$, we can easily find the other coefficient by employing \eqref{Bogoliubov:coefficients:main:text}, which gives 
\begin{align}\label{Bogoliubov:coefficients:constant:case:text}
\hat{\alpha}=&e^{i\,\hat{K}\,\tau}\,\left[\cos\left(\hat{\Lambda}\,\tau\right)-i\,\frac{\hat{K}}{\hat{\Lambda}}\,\sin\left(\hat{\Lambda}\,\tau\right)\right]\nonumber\\
\hat{\beta}=&-2\,i\,\tilde{\chi}_2\,\frac{a^\dagger\hat a}{\hat{\Lambda}}\,e^{2\,i\,\phi_2}\,e^{i\,\hat{K}\,\tau}\,\sin\left(\hat{\Lambda}\,\tau\right).
\end{align}
It is immediate to check that $\hat{\alpha}\,\hat{\alpha}^\dag-\hat{\beta}\,\hat{\beta}^\dag=\mathds{1}$.

This allows us to obtain an analytical expression for the time evolution operator \eqref{U}. We first obtain $\hat{F}^{(2)}$, which reads \eqref{F2:general}, together with an expression for
 $\Re\left[\tilde{g}_1(\tau)\,\hat{E}_2\,(\hat{\alpha}^\dag+\hat{\beta}^\dag)\right]$ and $\Im\left[\tilde{g}_1(\tau)\,\hat{E}_2\,(\hat{\alpha}^\dag-\hat{\beta}^\dag)\right]$, whose full expressions can be found in \eqref{useful:expression:one}.

We do not need an analytical expression for the decoupled form of the operator $\hat{U}_{\mathrm{sq}}$ for this case, since we have found the Bogoliubov coefficients \eqref{Bogoliubov:coefficients:constant:case:text}.

This means that our time evolution operator \eqref{U} has the explicit expression \eqref{U:constant:cross:Kerr:case:appendix}, which can be specialised to any desired functional expression of the $\tilde{g}_1$ drive.

\section{Conclusions}\label{conclusions}
We have studied the time evolution induced by a class of Hamiltonians of two interacting quantum harmonic oscillators which include cross-Kerr interactions. We employed tools developed to obtain an analytical expression for the time-evolution operator, which in turn allowed us to compute explicitly quantities of interest. These include the number expectation value of the mechanical resonator and the mixedness of the state of the resonator. Our results are free from approximations and therefore encode the full nonlinear character of the interaction. 

These techniques, and the control gained by employing them, can have many applications. For example, similar decoupling techniques have been recently applied the the analysis of correlations within tripartite coupled bosonic system interacting with quadratic Hamiltonians \cite{Bruschi:Sabin:2017}, which prompted and facilitated a successful experiment investigating the resulting entanglement between the bosonic modes \cite{Lahteenmaki:Paraoanu:2016}.
Among other possible applications of our tools there is quantum control \cite{Dong:Petersen:2010}, understanding of the interplay between the linear and nonlinear character of quantum mechanical systems \cite{qvarfort2018enhanced} and extending current studies of hidden quantum correlations in existing electromechanical measurements \cite{Ockeloen-Korrpi:Damskagg:2018}.

Finally, it is important to note that our system is ideal and \textit{closed}, in the sense that we have assumed it to be isolated from the environment, without losses and without decoherence. We leave it to future work to include these important aspects in order to achieve a more realistic and concrete analysis of optomechanical physics driven by cross-Kerr interactions.

\section*{Acknowledgments}
We would like to thank Anja Metelmann, Sorin Paraoanu, Alessio Serafini and Myung-Joong Hwang for useful comments. D. E. B. would like to thank the University of Vienna and the Erwin Schr\"odinger Institute (ESI) for hospitality.

\bibliographystyle{apsrev4-1}
\bibliography{bibliography}

\onecolumngrid

\newpage
\appendix

\section{Decoupling of techniques for time evolution}\label{appendix:decoupling}
In this appendix, we outline the general decoupling techniques that we shall be using throughout this work to find a decoupled time-evolution operator generated by the Hamiltonian in \eqref{main:time:dependent:Hamiltonian:to:decouple:general}. 

\subsection{Decoupling for arbitrary Hamiltonians}
The time evolution operator $\hat{U}(t)$ induced by a Hamiltonian $\hat{H}(t)$ reads
\begin{equation}
\hat{U}(t)=\overset{\leftarrow}{\mathcal{T}}\,\exp\left[-\frac{i}{\hbar}\int_0^{t} dt'\,\mathcal{H}(t')\right].
\end{equation}
Any Hamiltonian can be cast in the form $\hat{H}=\sum_n \hbar\, g_n(t)\,\hat{G}_n$, where the $\hat{G}_n$ are time independent, Hermitian operators and the $g_n(t)$ are time dependent functions. The choice of $\hat{G}_n$ need not be unique.

It has been shown \cite{bruschi2013time} that it is always possible to obtain the decoupling
\begin{equation}\label{decoupled:time:evolution:operator:appendix}
\hat{U}(t)=\prod_n\hat{U}_n(t),
\end{equation}
where we have defined $\hat{U}_n:=\exp[-i\,F_n(t)\,\hat{G}_n]$ and the real, time-dependent functions $F_n(t)$.

The functions $F_n(t)$ are uniquely determined by the coupled, nonlinear, first order differential equations
\begin{align}\label{operator:differential:equations}
\frac{1}{\hbar}\hat{H}=\dot{F}_1\,\hat{G}_1+\dot{F}_2\,\hat{U}_1\,\hat{G}_2\,\hat{U}_1^{\dag}+\dot{F}_3\,\hat{U}_1\,\hat{U}_2\,\hat{G}_3\,\hat{U}_2^{\dag}\,\hat{U}_1^{\dag}+\dot{F}_4\,\hat{U}_1\,\hat{U}_2\,\hat{U}_3\,\hat{G}_4\,\hat{U}_3^{\dag}\,\hat{U}_2^{\dag}\,\hat{U}_1^{\dag}+\dots,
\end{align}
This is the general method we has been employed in this work, and in previous related one \cite{}.

Here, we find it convenient to consider the \emph{closed} finite $9$-dimensional Lie algebra generated by the following set of Hermitian basis operators
\begin{align}\label{basis:operator:Lie:algebra}
	 	&(\hat a^\dagger \hat a)^N \nonumber  & (\hat a^\dagger \hat a)^N \hat b^\dagger \hat b \nonumber\\
		 &(\hat a^\dagger \hat a)^N\,\hat{B}_+ & (\hat a^\dagger \hat a)^N\,\hat{B}_-\nonumber\\
	&(\hat a^\dagger \hat a)^N \hat{B}^{(2)}_+ & (\hat a^\dagger \hat a)^N \hat{B}^{(2)}_-,  
\end{align}
which are the maximal extension of those that generate the Hamiltonian \eqref{main:time:dependent:Hamiltonian:to:decouple:general}. Note that there are an infinite of these operators, i.e., there are operators for all $N\in\mathbb{N}$. 

\subsection{Decoupling for quadratic Hamiltonians}
If the Hamiltonian is quadratic in the mode operators the techniques described in the previous subsection have a more powerful representation. Here we proceed to describe these techiques.

\subsection{Continuous variables and Covariance Matrix formalism}\label{tools:cmf}
In quantum mechanics, the initial state $\hat{\rho}_\textrm{I}$ of a system of $N$ bosonic modes with operators $\{\hat{a}_n,\hat{a}^{\dag}_n\}$ evolves to a final state $\hat{\rho}_f$ through the standard Heisenberg equation $\hat{\rho}_f=\hat{U}^{\dag}\,\hat{\rho}_\textrm{I}\,\hat{U}$, where $U$ implements the transformation of interest, such as time evolution. If the state $\hat{\rho}$ is Gaussian and the Hamiltonian $H$ is quadratic in the operators, it is convenient to introduce the vector $\hat{\mathbb{X}}=(\hat{a}_1,\ldots,\hat{a}_N,\hat{a}^{\dag}_1,\ldots,\hat{a}^{\dag}_N)^{Tp}$, the vector of first moments $d:=\langle\hat{\mathbb{X}}\rangle$ and the covariance matrix $\boldsymbol{\sigma}$ defined by $\sigma_{nm}:=\langle\{\hat{X}_n,\hat{X}^{\dag}_m\}\rangle-2\langle\hat{X}_n\rangle\langle\hat{X}_m^{\dag}\rangle$, where $\{\cdot,\cdot\}$ stands for anticommutator and all expectation values of an operator $\hat{\mathcal{A}}$ are defined by $\langle \hat{\mathcal{A}}\rangle:=\text{Tr}(\hat{\mathcal{A}}\,\hat{\rho})$. In this language, the canonical commutation relations read $[\hat{X}_n,\hat{X}_m^{\dag}]=i\,\Omega_{nm}$, where the $2N\times2N$ matrix $\boldsymbol{\Omega}$ is known as the symplectic form \cite{adesso2014continuous}. We then notice that, while arbitrary states of bosonic modes are, in general,  characterised by an infinite amount of degrees of freedom, a Gaussian state is uniquely determined by its first and second moments, $d_n$ and $\sigma_{nm}$ respectively \cite{adesso2014continuous}. Furthermore, quadratic (i.e., linear) unitary transformations, such as Bogoliubov transformations, preserve the Gaussian character of the Gaussian state and can always be represented by a $2N\times2N$ symplectic matrix $\boldsymbol{S}$ that preserves the symplectic form, i.e., $\boldsymbol{S}^{\dag}\,\boldsymbol{\Omega}\,\boldsymbol{S}=\boldsymbol{\Omega}$. 

The Heisenberg equation can be translated in this language to the simple equation $\boldsymbol{\sigma}_f=\boldsymbol{S}^{\dag}\,\boldsymbol{\sigma}_\textrm{I}\,\boldsymbol{S}$ for the second moments, and $\boldsymbol{r}_f = \boldsymbol{S} \, \boldsymbol{r}_\textrm{I} $ for the first moments, which shifts the problem of usually untreatable operator algebra to simple $2N\times2N$ matrix multiplication.  
Finally, Williamson's theorem guarantees that any $2N\times2N$ hermitian matrix, such as the covariance matrix $\boldsymbol{\sigma}$, can be decomposed as $\boldsymbol{\sigma}=\boldsymbol{S}^{\dag}\,\boldsymbol{\nu}_{\oplus}\,\boldsymbol{S}$, where $\boldsymbol{S}$ is an appropriate symplectic matrix, the diagonal matrix $\boldsymbol{\nu}_{\oplus}=\textrm{diag}(\nu_1,\dots,\nu_N,\nu_1,\dots,\nu_N)$ is known as the Williamson form of the state and $\nu_n:=\coth(\frac{2\,\hbar\,\omega_n}{k_B\,T})\geq1$ are the symplectic eigenvalues of the state \cite{williamson1936algebraic}.

Williamson's form $\boldsymbol{\nu}_{\oplus}$ contains information about the local and global mixedness of the state of the system \cite{adesso2014continuous}. The state is pure when $\nu_n = 1$ for all $n $
and is mixed otherwise. As an example, the thermal state $\boldsymbol{\sigma}_{th}$ of a $N$-mode bosonic system is simply given by its Williamson form, i.e., $\boldsymbol{\sigma}_{th}=\boldsymbol{\nu}_{\oplus}$.

The time evolution operator has a symplectic representation $\boldsymbol{S}(t)=\overset{\leftarrow}{\mathcal{T}}\exp[\boldsymbol{\Omega}\,\int_0^\tau\boldsymbol{H}]$, where $\hat{H}=\frac{1}{2}\,\mathbb{X}^{\dag}\,\boldsymbol{G}_n\,\mathbb{X}$, and the decoupled ansatz \eqref{decoupled:time:evolution:operator:appendix} has the form 
\begin{equation}\label{general:symplectic:decomposition}
\boldsymbol{S}=\prod_{n=1}^{N\,(2N+1)}\boldsymbol{S}_n,
\end{equation}
where we have introduced  $\boldsymbol{S}_n:=\exp[F_n(t)\,\boldsymbol{\Omega}\,\boldsymbol{S}_n]$, the matrices $\boldsymbol{S}_n$ through $\hat{G}_n=\frac{1}{2}\,\mathbb{X}^{\dag}\,\boldsymbol{G}_n\,\mathbb{X}$, and the real time-dependent functions $F_n(t)$ that are the \emph{same} as they would be obtained with the technique above.

The real, time dependent functions $F_n(\tau)$ can be obtained by solving the following system of coupled nonlinear first order differential equations
\begin{align}\label{matrix:differential:equations}
\frac{1}{\hbar}\boldsymbol{H}=&\dot{F}_1\,\boldsymbol{G}_1+\dot{F}_2\,\boldsymbol{S}_1^{\dag}\,\boldsymbol{G}_2\,\boldsymbol{S}_1+\dot{F}_3\,\boldsymbol{S}_1^{\dag}\,\boldsymbol{S}_2^{\dag}\,\boldsymbol{G}_3\,\boldsymbol{S}_2\,\boldsymbol{S}_1+\dot{F}_4\,\boldsymbol{S}_1^{\dag}\,\boldsymbol{S}_2^{\dag}\,\boldsymbol{S}_3^{\dag}\,\boldsymbol{G}_4\,\boldsymbol{S}_3\,\boldsymbol{S}_2\,\boldsymbol{S}_1+\dots.
\end{align}
This is the matrix version of the operator differential equations \eqref{operator:differential:equations} for quadratic Hamiltonians, which reduces the problem of operator algebra to matrix multiplication.

\section{Decoupling of the squeezing Hamiltonian}\label{appendix:Hamiltonian:nonlinear:decoupling}
Here we employ the techniques of Appendix \eqref{appendix:decoupling} to decouple the operator $\hat{U}_{\mathrm{sq}}=\overset{\leftarrow}{\mathcal{T}}\exp\left[-\frac{i}{\hbar}\int_0^\tau\,d\tau'\,\hat{H}_{\mathrm{sq}}(\tau')\right]$ induced by the squeezing Hamiltonian $\hat{H}_{\mathrm{sq}}(\tau)$, which we reprint here
\begin{align}
\hat{H}_{\mathrm{sq}}(\tau):=a^\dagger\hat a\,\left[\tilde{g}^{(+)}_2(\tau)\,\cos(2\,\hat{\theta}_2)-\tilde{g}^{(-)}_2(\tau)\,\sin(2\,\hat{\theta}_2)\right]\,\hat{B}^{(2)}_+
+a^\dagger\hat a\,\left[\tilde{g}^{(+)}_2(\tau)\,\sin(2\,\hat{\theta}_2)+\tilde{g}^{(-)}_2(\tau)\,\cos(2\,\hat{\theta}_2)\right]\,\hat{B}^{(2)}_-.
\end{align}

\subsection{Preliminaries}
We start by defining $\mathbb{X}:=(\hat b, \hat b^\dag)^{\mathrm{T}}$ an noting that we can write
\begin{align} \label{eq:Bogoliubov:transform}
\mathbb{X}'
=
\hat{U}_{\mathrm{sq}}^\dag
\,\mathbb{X}\,
\hat{U}_{\mathrm{sq}}
=
\begin{pmatrix}
\hat{U}_{\mathrm{sq}}^\dag\,\hat b\,\hat{U}_{\mathrm{sq}} \\
\hat{U}_{\mathrm{sq}}^\dag\,\hat b^\dag\,\hat{U}_{\mathrm{sq}}
\end{pmatrix}
=
\boldsymbol{S}_{\mathrm{sq}}(\tau)\,
\mathbb{X},
\end{align}
where the $2\times2$ symplectic matrix $\boldsymbol{S}_{\mathrm{sq}}(\tau)$ is the symplectic representation of $\hat{H}_{\mathrm{sq}}(\tau)$ and satisfies $\boldsymbol{S}_{\mathrm{sq}}^\dag(\tau)\,\boldsymbol{\Omega}\,\boldsymbol{S}_{\mathrm{sq}}(\tau)=\boldsymbol{\Omega}$. Here $\boldsymbol{\Omega}=\text{diag}(-i,i)$ is the symplectic form. The matrix $\boldsymbol{S}_{\mathrm{sq}}(\tau)$ therefore has the expression $\boldsymbol{S}_{\mathrm{sq}}(\tau)=\overset{\leftarrow}{\mathcal{T}}\,\exp[\boldsymbol{\Omega}\,\int_0^\tau\,d\tau'\,\boldsymbol{H}_{\mathrm{sq}}(\tau')]$. In this case we have 
\begin{align}
\hat {H}_{\mathrm{sq}}
=
\frac{1}{2}\,\mathbb{X}^\dag
\boldsymbol{H}_{\mathrm{sq}}
\mathbb{X},
\,\,\,\,\,\,\,\,\,\,\,\,\,\,\,\,\,\,\text{with}\,\,\,\,\,\,
\tilde{\boldsymbol{H}}_{\mathrm{sq}}
=2\,\hat{a}^\dagger\hat a\,
\begin{pmatrix}
0 & \tilde{g}_2(\tau)\,e^{2\,i\,\hat{\theta}_2}\\
\tilde{g}^*_2(\tau)\,e^{-2\,i\,\hat{\theta}_2} & 0 
\end{pmatrix},
\end{align}
where we have defined $\tilde{g}_2(\tau)=\tilde{\chi}_2(\tau)\,\exp[2\,i\,\phi_2]:=\tilde{g}^{(+)}_2(\tau)+i\,\tilde{g}^{(-)}_2(\tau)$, and therefore we have $\tilde{\rho}(\tau)=\sqrt{\tilde{g}^{(+)2}_2(\tau)+\tilde{g}^{(-)2}_2(\tau)}\geq0$ and $\tan(2\,\phi_2)=\tilde{g}^{(-)}_2/\tilde{g}^{(+)}_2$.

Therefore, through simple algebra we obtain
\begin{align}\label{matrix:time:ordered:exponential}
\boldsymbol{S}_{\mathrm{sq}}(\tau)=\overset{\leftarrow}{\mathcal{T}}\,\exp\left[-2\,i\,\int_0^\tau\,d\tau'\,\tilde{\chi}_2(\tau')\,\hat a^\dagger\hat a\,
\begin{pmatrix}
0 & \,e^{2\,i\,(\phi_2+\hat{\theta}_2)}\\
-e^{-2\,i\,(\phi_2+\hat{\theta}_2)} & 0 
\end{pmatrix}
\right].
\end{align}

\subsection{Solving the matrix time-ordered exponential}
Here we wish to find a formal expression for \eqref{matrix:time:ordered:exponential}.
We start by noticing that, if we wrote down the time ordered exponential we would be able to write
\begin{align}\label{initial:useful:conditions}
\overset{\leftarrow}{\mathcal{T}}\,\exp\left[-i\,\hat a^\dagger\hat a \,\int_0^\tau\,d\tau'\,\boldsymbol{K}(\tau')\right]:=\overset{\leftarrow}{\mathcal{T}}\,\exp\left[-i\,2\,\hat{a}^\dagger\hat a\,\int_0^\tau\,d\tau'\,\tilde{\chi}_2(\tau')\,
\begin{pmatrix}
0 & \,e^{2\,i\,(\phi_2+\hat{\theta}_2)}\\
-e^{-2\,i\,(\phi_2+\hat{\theta}_2)} & 0 
\end{pmatrix}
\right]=\boldsymbol{P}-i\,a^\dagger\hat a\,\int_0^\tau d\tau'\,\boldsymbol{K}\,\boldsymbol{P},
\end{align}
where the matrix $\boldsymbol{K}$ is defined as
\begin{align}\label{matrix:time:ordered:exponential:else}
\boldsymbol{K}:=2\,\tilde{\chi}_2(\tau)\,
\begin{pmatrix}
0 & \,e^{2\,i\,(\phi_2+\hat{\theta}_2)}\\
-e^{-2\,i\,(\phi_2+\hat{\theta}_2)} & 0 
\end{pmatrix},
\end{align}
and the \emph{diagonal} matrix $\boldsymbol{P}$ is the object that we need to compute now. Note that it is straightforward to check that $\boldsymbol{P}$ is diagonal.

We use the fact that 
\begin{align}
\frac{d}{d\tau}\overset{\leftarrow}{\mathcal{T}}\,\exp\left[-i\,\hat a^\dagger\hat a \,\int_0^\tau\,d\tau'\,\boldsymbol{K}(\tau')\right]
=
-i\,\hat a^\dagger\hat a \,\boldsymbol{K}\,
\overset{\leftarrow}{\mathcal{T}}\,\exp\left[-i\,\hat a^\dagger\hat a \,\int_0^\tau\,d\tau'\,\boldsymbol{K}(\tau')\right]
\end{align}
to find the equation
\begin{align}
-(a^\dagger\hat a)^2\,\boldsymbol{K}\,\int_0^\tau d\tau'\,\boldsymbol{K}\,\boldsymbol{P}=\dot{\boldsymbol{P}}.
\end{align}
Since $\boldsymbol{K}$ is invertible (for all cases \textit{except} when $\tilde{\chi}_2(\tau)$=0, which implies $\boldsymbol{P}=\mathds{1}$), we can manipulate this equation and obtain, after some algebra,
\begin{align}\label{differential:equation:for:matrix}
\ddot{\boldsymbol{P}}-
\begin{pmatrix}
\dot{\tilde{\chi}}_2(\tau)/\tilde{\chi}_2(\tau)+2\,i\,\frac{d}{d\tau}(\phi_2+\hat{\theta}_2) &  0\\
0 & \dot{\tilde{\rho}}_2(\tau)/\tilde{\chi}_2(\tau)-2\,i\,\frac{d}{d\tau}(\phi_2+\hat{\theta}_2) 
\end{pmatrix}
\dot{\boldsymbol{P}}
-4\,\tilde{\chi}_2^2(\tau)\,(a^\dagger\hat a)^2\,\boldsymbol{P}=0.
\end{align}
We can now solve the four differential equations contained in \eqref{differential:equation:for:matrix}, two of which are trivial and read $P_{12}=P_{21}=0$, which read
\begin{align}\label{differential:equation:written:down}
\ddot{P}_{11}-\left(\dot{\tilde{\rho}}_2(\tau)/\tilde{\chi}_2(\tau)+2\,i\,\frac{d}{d\tau}(\phi_2+\hat{\theta}_2)\right)\,\dot{P}_{11}-4\,\tilde{\chi}_2^2(\tau)\,(a^\dagger\hat a)^2\,P_{11}=&0\nonumber\\
\ddot{P}_{22}-\left(\dot{\tilde{\rho}}_2(\tau)/\tilde{\chi}_2(\tau)-2\,i\,\frac{d}{d\tau}(\phi_2+\hat{\theta}_2)\right)\,\dot{P}_{22}-4\,\tilde{\chi}_2^2(\tau)\,(a^\dagger\hat a)^2\,P_{22}=&0.
\end{align}
The differential equations \eqref{differential:equation:written:down} do not admit an explicit solution in general. However, they can be integrated numerically when an explicit form of $\tilde{g}_2(\tau)$ and $\tilde{g}'(\tau)$ are given.

These differential equations have to be supplemented by initial conditions. We note that, since the left hand side of \eqref{initial:useful:conditions} is the identity matrix for $\tau=0$, we have that $\boldsymbol{P}(0)=0$ which implies $\hat{p}_{11}(0)=P_{22}(0)=1$. In addition, taking the time derivative of both sides of \eqref{initial:useful:conditions} and setting $t=0$ it is easy to check that this implies $\dot{P}_{11}(0)=\dot{P}_{22}(0)=0$.

Given that there differential equations are valid only when $\tilde{g}_2(\tau)\neq0$, it is convenient to introduce the functions $\hat{p}_{11}(y)$ and $\hat{p}_{22}(y)$ defined as
\begin{align}\label{useful:functions}
\hat{P}_{11}(\tau)=\hat{p}_{11}(y):=&\hat{p}_{11}\left(2\,\int_0^\tau\,d\tau'\tilde{\chi}_2(\tau')\right)\nonumber\\
\hat{P}_{22}(\tau)=\hat{p}_{22}(y):=&\hat{p}_{22}\left(2\,\int_0^\tau\,d\tau'\tilde{\chi}_2(\tau')\right)\nonumber\\
\end{align}
where we have introduced $y(\tau):=2\,\int_0^\tau\,d\tau'\tilde{\chi}_2(\tau)$.
In terms of these functions we have that the differential equations \eqref{differential:equation:written:down} read
\begin{align}\label{differential:equation:written:down:simpler}
\tilde{\chi}_2(\tau)\,\ddot{\hat{p}}_{11}-i\,\frac{d}{d\tau}(\phi_2+\hat{\theta}_2)\,\dot{\hat{p}}_{11}-\tilde{\chi}_2(\tau)\,(a^\dagger\hat a)^2\,\hat{p}_{11}=&0\nonumber\\
\tilde{\chi}_2(\tau)\,\ddot{\hat{p}}_{22}+i\,\frac{d}{d\tau}(\phi_2+\hat{\theta}_2)\,\dot{\hat{p}}_{22}-\tilde{\chi}_2(\tau)\,(a^\dagger\hat a)^2\,p_{22}=&0,
\end{align}
where the derivative stands for derivative with respect to the variable $y$.

This has allowed us to find
\begin{align}
\boldsymbol{S}_{\mathrm{sq}}(\tau)=&\boldsymbol{P}-i\,a^\dagger\hat a\,\int_0^\tau d\tau'\,\boldsymbol{K}\,\boldsymbol{P}=
\begin{pmatrix}
\hat{p}_{11} &  -2\,i\,a^\dagger\hat a\, \int_0^\tau d\tau'\,\tilde{\chi}_2(\tau')\,e^{2\,i\,(\phi_2+\hat{\theta}_2)}\,\hat{p}_{22}\\
2\,i\,a^\dagger\hat a\, \int_0^\tau d\tau'\,\tilde{\chi}_2(\tau')\,e^{-2\,i\,(\phi_2+\hat{\theta}_2)}\,\hat{p}_{11} & \hat{p}_{22} 
\end{pmatrix}.
\end{align}
Given that we know that, in general, one has
\begin{align}
\boldsymbol{S}_{\mathrm{sq}}(\tau)=&
\begin{pmatrix}
\hat{\alpha} &  \hat{\beta}\\
\hat{\beta}^\dag &\hat{\alpha}^\dag
\end{pmatrix},
\end{align}
this immediately allows us to identify the Bogoliubov coefficients as
\begin{align}
\hat{\alpha}=&\hat{p}_{11} \nonumber\\
\hat{\beta}=&-2\,i\,\hat{a}^\dagger\hat a\, \int_0^\tau d\tau'\,\tilde{\chi}_2(\tau')\,e^{2\,i\,(\phi_2+\hat{\theta}_2)}\,\hat{p}^\dag_{11},
\end{align}
with the auxiliary consistency condition $\hat{p}_{22}=\hat{p}^\dag_{11}$. This condition follows also from \eqref{differential:equation:written:down}.

\subsection{Useful expressions}\label{appendix:subsection:useful:expressions}
Here we present a list of useful expressions that are too cumbersome to appear in the main text, but include key steps for the obtainment of the final results.

We start by presenting the expressions for $\Re\left[\tilde{g}_1(\tau)\,\hat{E}_2\,(\hat{\alpha}^\dag+\hat{\beta}^\dag)\right]$ and $\Im\left[\tilde{g}_1(\tau)\,\hat{E}_2\,(\hat{\alpha}^\dag-\hat{\beta}^\dag)\right]$ for the case of constant couplings $\tilde{\chi}_2$ and $\tilde{g}^\prime_2$, which read
\begin{align}\label{useful:expression:one}
\Re\left[\tilde{g}_1(\tau)\,\hat{E}_2\,(\hat{\alpha}^\dag+\hat{\beta}^\dag)\right]=&\tilde{g}_1^{(+)}\,\left[\cos\left(\hat{\Lambda}\,\tau\right)+2\,\tilde{\chi}_2\,\frac{a^\dagger\hat a}{\hat{\Lambda}}\,\sin(2\,\phi_2)\,\sin\left(\hat{\Lambda}\,\tau\right)\right]-\tilde{g}_1^{(-)}\,\left[\frac{\hat{K}}{\hat{\Lambda}}+2\,\tilde{\chi}_2\,\frac{a^\dagger\hat a}{\hat{\Lambda}}\,\cos(2\,\phi_2)\right]\,\sin\left(\hat{\Lambda}\,\tau\right)\nonumber\\
\Im\left[\tilde{g}_1(\tau)\,\hat{E}_2\,(\hat{\alpha}^\dag-\hat{\beta}^\dag)\right]=&\tilde{g}_1^{(+)}\,\left[\frac{\hat{K}}{\hat{\Lambda}}-2\,\tilde{\chi}_2\,\frac{a^\dagger\hat a}{\hat{\Lambda}}\,\cos(2\,\phi_2)\right]\,\sin\left(\hat{\Lambda}\,\tau\right)+\tilde{g}_1^{(-)}\,\left[\cos\left(\hat{\Lambda}\,\tau\right)-2\,\tilde{\chi}_2\,\frac{a^\dagger\hat a}{\hat{\Lambda}}\,\sin(2\,\phi_2)\,\sin\left(\hat{\Lambda}\,\tau\right)\right].
\end{align}
This allows us to find
\begin{align}\label{U:constant:cross:Kerr:case:appendix}
\hat U(t):=&e^{-i\,\int_0^\tau\,d\tau'\tilde{\omega}_\textrm{c}(\tau') \hat{a}^\dag\,\hat{a}+i\,\hat{F}^{(2)}\,(\hat{a}^\dagger\hat a\,)^2}\, e^{-i\,\hat{\theta}_2\,\hat{b}^\dag\,\hat{b}}\,\hat{U}_{\mathrm{sq}}\,e^{-i\,\hat{a}^\dagger\hat a\,\int_0^\tau\,d\tau'\,\tilde{g}_1^{(+)}(\tau)\,\left[\cos\left(\hat{\Lambda}\,\tau'\right)+2\,\tilde{\chi}_2\,\frac{a^\dagger\hat a}{\hat{\Lambda}}\,\sin(2\,\phi_2)\,\sin\left(\hat{\Lambda}\,\tau'\right)\right]\,\hat{B}_+}\nonumber\\
&\times e^{i\,\hat{a}^\dagger\hat a\,\left[\frac{\hat{K}}{\hat{\Lambda}}+2\,\tilde{\chi}_2\,\frac{a^\dagger\hat a}{\hat{\Lambda}}\,\cos(2\,\phi_2)\right]\,\int_0^\tau\,d\tau'\,\tilde{g}_1^{(-)}(\tau')\,\sin\left(\hat{\Lambda}\,\tau'\right)\,\hat{B}_+}\,e^{i\,\hat{a}^\dagger\hat a\,\left[\frac{\hat{K}}{\hat{\Lambda}}-2\,\tilde{\chi}_2\,\frac{a^\dagger\hat a}{\hat{\Lambda}}\,\cos(2\,\phi_2)\right]\,\int_0^\tau\,d\tau'\,\tilde{g}_1^{(+)}(\tau')\sin\left(\hat{\Lambda}\,\tau'\right)\,\hat{B}_-}\nonumber\\
&\times e^{i\,\hat{a}^\dagger\hat a\,\int_0^\tau\,d\tau'\,\tilde{g}_1^{(-)}(\tau')\,\left[\cos\left(\hat{\Lambda}\,\tau'\right)-2\,\tilde{\chi}_2\,\frac{a^\dagger\hat a}{\hat{\Lambda}}\,\sin(2\,\phi_2)\,\sin\left(\hat{\Lambda}\,\tau'\right)\right]\,\hat{B}_-}.
\end{align}

\section{Properties of the final state}\label{appendix:section:final:state}
The final state $\hat{\rho}(\tau)$ can be partially computed analytically, given the expressions for our initial state \eqref{initial:state:two} and the time-evolution operator \eqref{U}, and reads
\begin{align}\label{final:state}
\hat{\rho}(\tau)=&e^{-|\mu|^2}\sum_{n,m,p}\frac{\tanh^{2\,p} r_T}{\cosh^2r_T}\frac{\mu^n\,\mu^{*m}}{\sqrt{n!}\sqrt{m!}}\,\hat U(t)|n\rangle|p\rangle\langle m|\langle p|\hat U^\dag(t),
\end{align}
which can be made slightly more explicit as
\begin{align}\label{final:state:more:explicit}
\hat{\rho}(\tau)=&e^{-|\mu|^2}\sum_{n,m,p}\frac{\tanh^{2\,p} r_T}{\cosh^2r_T}\frac{\mu^n\,\mu^{*m}}{\sqrt{n!}\sqrt{m!}}\,
e^{-i\,\int_0^\tau\,d\tau'\tilde{\omega}_\textrm{c}(\tau') (n-m)+i\,(\hat{F}_n^{(2)}\,n^2-\hat{F}_m^{(2)}\,m^2)}\nonumber\\
&\times e^{-i\,\hat{\theta}_{2,n}\,\hat{b}^\dag\,\hat{b}}\,\hat{U}_{\mathrm{sq},n}\,e^{-i\,\int_0^\tau\,d\tau'\,n\,\Re\left[\tilde{g}_1(\tau')\,\hat{E}_2\,(\hat{\alpha}^\dag+\hat{\beta}^\dag)\right]_n\,\hat{B}_+}\,e^{i\,\int_0^\tau\,d\tau'\,n\,\Im\left[\tilde{g}_1(\tau')\,\hat{E}_2\,(\hat{\alpha}^\dag-\hat{\beta}^\dag)\right]_n\,\hat{B}_-}\nonumber\\
&|n\rangle|p\rangle\langle m|\langle p|e^{-i\,\int_0^\tau\,d\tau'\,m\,\Im\left[\tilde{g}_1(\tau')\,\hat{E}_2\,(\hat{\alpha}^\dag-\hat{\beta}^\dag)\right]_m\,\hat{B}_-}\,e^{i\,\int_0^\tau\,d\tau'\,m\,\Re\left[\tilde{g}_1(\tau')\,\hat{E}_2\,(\hat{\alpha}^\dag+\hat{\beta}^\dag)\right]_m\,\hat{B}_+}\hat{U}^\dag_{\mathrm{sq},m}\,e^{i\,\hat{\theta}_{2,m}\,\hat{b}^\dag\,\hat{b}}.
\end{align}
In the expression \eqref{final:state:more:explicit}, the subscript $n$ and $m$ mean that the respective quantities are evaluated for $\hat{a}^\dag\hat{a}\rightarrow n,m$.

\subsection{Reduced final state of the mechanical oscillator}\label{appendix:section:final:state:reduced}
We can compute the reduced the final reduced state $\hat{\rho}_{\textrm{m}}(\tau)$ of the mechanical resonator. We can be obtain it as $\hat{\rho}_{\textrm{m}}(\tau):=\textrm{Tr}_{\textrm{c}}(\hat{\rho}(\tau))$, which we can compute using \eqref{final:state:more:explicit}. We find
\begin{align}\label{reduced:final:state:more:explicit}
\hat{\rho}_{\textrm{m}}(\tau)=&e^{-|\mu|^2}\sum_{n,p}\frac{\tanh^{2\,p} r_T}{\cosh^2r_T}\frac{|\mu|^{2\,n}}{n!}\,
 e^{-i\,\hat{\theta}_{2,n}\,\hat{b}^\dag\,\hat{b}}\,\hat{U}_{\mathrm{sq},n}\,e^{-i\,\int_0^\tau\,d\tau'\,n\,\Re\left[\tilde{g}_1(\tau')\,\hat{E}_2\,(\hat{\alpha}^\dag+\hat{\beta}^\dag)\right]_n\,\hat{B}_+}\,e^{i\,\int_0^\tau\,d\tau'\,n\,\Im\left[\tilde{g}_1(\tau')\,\hat{E}_2\,(\hat{\alpha}^\dag-\hat{\beta}^\dag)\right]_n\,\hat{B}_-}\nonumber\\
&|p\rangle\langle p|e^{-i\,\int_0^\tau\,d\tau'\,n\,\Im\left[\tilde{g}_1(\tau')\,\hat{E}_2\,(\hat{\alpha}^\dag-\hat{\beta}^\dag)\right]_n\,\hat{B}_-}\,e^{i\,\int_0^\tau\,d\tau'\,n\,\Re\left[\tilde{g}_1(\tau')\,\hat{E}_2\,(\hat{\alpha}^\dag+\hat{\beta}^\dag)\right]_n\,\hat{B}_+}\hat{U}^\dag_{\mathrm{sq},n}\,e^{i\,\hat{\theta}_{2,n}\,\hat{b}^\dag\,\hat{b}}.
\end{align}
Note that here, again, the subscript $n$ means that we need to replace $n\rightarrow\hat{a}^\dag\,\hat{a}$.

\section{Mixedness in the final reduced state of the mechanical oscillator}\label{appendix:subsection:mixedness:computation}
Here we compute the mixedness of the reduced state of the mechanical oscillator. Given that any pure state $\hat{\rho}$ has the property that $\hat{\rho}^2=\hat{\rho}$, it follows that the mixedness of a state can be quantified by the \textit{linear entropy} $S_N$ defined as $S_N(\hat{\rho}):=1-\textrm{Tr}(\hat{\rho}^2)$.

In general, although the expression \eqref{reduced:final:state:more:explicit} gives us some insight on the reduced state of the mechanical mode, it is difficult to obtain the mixedness for arbitrary parameters of the Hamiltonian. We therefore proceed to specialise to interesting, yet general enough, scenarios below.

\subsection{Mixedness of the reduced state of the mechanical oscillator: no cross Kerr squeezing}\label{appendix:subsection:mixedness:Kerr:squeezing:computation}
In the case where $\tilde{g}_2=0$ we have $\hat{U}_{\mathrm{sq},n}=\hat{\alpha}=\mathds{1}$ and $\hat{\beta}=0$. Therefore, with some algebra we can show that the expression \eqref{reduced:final:state:more:explicit} simplifies to
\begin{align}\label{reduced:final:state:no:cross:Kerr:squeezing:explicit}
\hat{\rho}_{\textrm{m}}(\tau)=&e^{-|\mu|^2}\sum_{n,p}\frac{\tanh^{2\,p} r_T}{\cosh^2r_T}\frac{|\mu|^{2\,n}}{n!}\,
 e^{-i\,\hat{\theta}_2\,\hat{b}^\dag\,\hat{b}}\,e^{-i\,\int_0^\tau\,d\tau'\,n\,\left[\tilde{g}^*_1(\tau')\,e^{-i\,\hat{\theta}_{2,n}}\,\hat{b}^\dag+\tilde{g}_1(\tau')\,e^{i\,\hat{\theta}_{2,n}}\,\hat{b}\right]}|p\rangle\langle p|\nonumber\\
 &\times e^{i\,\int_0^\tau\,d\tau'\,n\,\left[\tilde{g}^*_1(\tau')\,e^{-i\,\hat{\theta}_{2,n}}\,\hat{b}^\dag+\tilde{g}_1(\tau')\,e^{i\,\hat{\theta}_{2,n}}\,\hat{b}\right]}\,e^{i\,\hat{\theta}_2\,\hat{b}^\dag\,\hat{b}}.
\end{align}
Therefore, we can easily show that the linear entropy $S_N(\hat{\rho}_{\textrm{m}}(\tau))$ associated to this state reads
\begin{align}\label{intermediary:formula}
S_N(\hat{\rho}_{\textrm{m}}(\tau))=&1-e^{-2\,|\mu|^2}\sum_{n,n',p,p'}\frac{\tanh^{2\,p+2\,p'} r_T}{\cosh^4r_T}\frac{|\mu|^{2\,n}}{n!}\,\frac{|\mu|^{2\,n'}}{n'!}\,\left|\langle p|\hat{D}_{nn'}(\tau)|p'\rangle \right|^2,
\end{align}
where we have introduced
\begin{align}
\hat{D}_{nn'}(\tau):=& e^{i\,\int_0^\tau\,d\tau'\,n\,\left[\tilde{g}^*_1(\tau')\,e^{-i\,\hat{\theta}_{2,n}}\,\hat{b}^\dag+\tilde{g}_1(\tau')\,e^{i\,\hat{\theta}_{2,n}}\,\hat{b}\right]}
\, e^{-i\,\int_0^\tau\,d\tau'\,n'\,\left[\tilde{g}^*_1(\tau')\,e^{-i\,\hat{\theta}_{2,n'}}\,\hat{b}^\dag+\tilde{g}_1(\tau')\,e^{i\,\hat{\theta}_{2,n'}}\,\hat{b}\right]}\nonumber\\
=&e^{i\,\psi_{nn'}}\,e^{i\,\Delta_{nn'}\,\hat{b}}\,e^{i\,\Delta_{nn'}\,\hat{b}^\dag},
\end{align}
and defined $\Delta_{nn'}:=\int_0^\tau\,d\tau'\,\tilde{g}_1(\tau')\bigl(n\,e^{i\,\tau'+2\,i\,n\,\int_0^{\tau'}\,d\tau''\,\tilde{g}_2^\prime(\tau'')}-n'\,e^{i\,\tau'+2\,i\,n'\,\int_0^{\tau'}\,d\tau''\,\tilde{g}_2^\prime(\tau'')}\bigr)$. Note that the exact form of $e^{i\,\psi_{nn'}}$ does not matter since it cancels out in \eqref{intermediary:formula}.

It is also easy to check that $S_N(\hat{\rho}_{\textrm{m}}(\tau))$ can be also written as
\begin{align}\label{linear:entropy:to:compute:appendix}
S_N(\hat{\rho}_{\textrm{m}}(\tau))=&1-2\,e^{-2\,|\mu|^2}\sum_{n,n',p=0}\sum_{k=1}\frac{\tanh^{4\,p+2\,k} r_T}{\cosh^4r_T}\frac{|\mu|^{2\,n}}{n!}\,\frac{|\mu|^{2\,n'}}{n'!}\,\left|\langle p|\hat{D}_{nn'}(\tau)|p+k\rangle \right|^2\nonumber\\
&-e^{-2\,|\mu|^2}\sum_{n,n',p=0}\frac{\tanh^{4\,p} r_T}{\cosh^4r_T}\frac{|\mu|^{2\,n}}{n!}\,\frac{|\mu|^{2\,n'}}{n'!}\, \left|\langle p|\hat{D}_{nn'}(\tau)|p\rangle \right|^2.
\end{align}
Now, we can follow the same procedure introduced in \cite{Bruschi:2018} and compute \eqref{linear:entropy:to:compute:appendix} explicitly. We find
\begin{align}\label{linear:entropy:final:appendix}
S_N(\hat{\rho}_{\textrm{m}}(\tau))=&1-e^{-2\,|\mu|^2}\sum_{n,n'}\frac{|\mu|^{2\,n+2\,n'}}{n!\,n'!}\,\frac{\exp\left[-\frac{1}{\cosh r_T}\,\left|\Delta_{nn'}\right|^2\right]}{\cosh r_T}.
\end{align}
We note that, as expected, the result \eqref{linear:entropy:final:appendix} reduces to the existing one in the literature for $\tilde{g}_2^\prime(\tau)=0$, see \cite{Bruschi:2018}.

\end{document}